\documentclass[10pt]{article}
\usepackage{amsmath}
\usepackage{amssymb}
\usepackage{graphicx}

\begin{document}

\setcounter{figure}{0}
\setcounter{table}{0}
\setcounter{footnote}{0}
\setcounter{equation}{0}

\vspace*{0.5cm}

\noindent {\Large ON THE GALACTIC ABERRATION CONSTANT}
\vspace*{0.7cm}

\noindent\hspace*{1.5cm} Z. MALKIN\\
\noindent\hspace*{1.5cm} Pulkovo Observatory, St. Petersburg, Russia\\
\noindent\hspace*{1.5cm} St. Petersburg State University, St. Petersburg, Russia\\
\noindent\hspace*{1.5cm} e-mail: malkin@gao.spb.ru\\

\vspace*{0.5cm}

\noindent {\large ABSTRACT.}
In this work, we analyzed all available determinations of the Galactic rotation parameters $R_0$ and $\Omega_0$ made during last 10 years to derive
the most probable value of the Galactic aberration constant $A = R_0 \Omega_0^2/c$.
We used several statistical methods to obtain reliable estimates of $R_0$ and $\Omega_0$ and their realistic errors.
In result, we have obtained the value of $A = 5.0 \pm 0.3$~$\mu$as/yr as the current best estimate of the GA constant.
We suggest that the proposed value of the GA constant can be safely used in practice during coming years.

\vspace*{1cm}

\noindent {\large 1. INTRODUCTION}

\smallskip

Galactic aberration (GA) is a small effect in proper motion of about 5~$\mu$as/yr already noticeable in VLBI and other highly-accurate astrometric
observations.
However accounting for this effect during data processing faces difficulty caused by the uncertainty in the GA constant $A = R_0 \Omega_0^2/c$,
where $R_0$ is the Galactocentric distance of the Sun, $\Omega_0$ is the angular velocity of circular rotation of the Sun around the Galactic center,
$c$ is speed of light.

The value of the GA constant can be derived either using the stellar astronomy methods or VLBI observations of the extragalactic radio sources.
It seems that the former provide more accurate results, while the latter are still somewhat contradictory.
So, we use the results of the observations of Galactic objects to improve $A$.
Our previous estimate of the GA constant (Malkin 2011) yields the values of $R_0$ = 8.2~kpc, $\Omega_0$ = 29.5~km~s$^{-1}$~kpc$^{-1}$,
and $A$ = 5.02$~\mu$as/yr.
This work is performed to check and improve if necessary this estimate taking into account more recent measurements of the Galactic rotation parameters.

\vspace*{0.7cm}

\noindent {\large 2. DERIVING THE BEST VALUE OF THE GA CONSTANT}

\smallskip

In this work, we have used 35 $R_0$ measurements and 30 $\Omega_0$ measurements made during last 10 years.
They are listed in Table~\ref{tab:data}.
We consider the results obtained during last 5 years as the most reliable, especially for $R_0$ estimates, for which the direct methods,
such as measurements of the parallax or stellar orbits around the massive black hole, become routine starting from 2008.
So, the results published in 2008--2013 were used to derive the final estimate of the GA constant.
The results of 2003--2007 were processed for control of its stability.

We have applied several statistical techniques mostly used in physics and metrology to these data, as described in Malkin (2012, 2013).
Results of computation are presented in Table~\ref{tab:result}.
The first line corresponds to the best current estimates of the GA constant, in our opinion.
The second result obtained by using only direct $R_0$ measurements is practically the same.
It shows that the results of the direct determinations of $R_0$ does not substantially differ (in average) from other estimates.
The results obtained with all measurements of the Galactic rotation parameters made during last 10 years are given in the third line.
We think it is less reliable than the first two ones.
However, it allows one to get an impression about the stability of the GA constant in time.

For comparison, using the standard weighted mean estimate we have got for the main variant (data interval of 2008--2013 and
using all $R_0$ measurements) corresponding to the first line of Table~\ref{tab:result}
$R_0 = 8.03 \pm 0.06$ kpc, $\Omega_0 = 29.23 \pm 0.19$ km s$^{-1}$ kpc$^{-1}$, $A = 4.83 \pm 0.07$~$\mu$as/yr.
Precision of these estimates seems to be too optimistic.
Using combined estimate from different statistical techniques as suggested by Malkin (2012) provides more reliable $A$ estimate
with a realistic uncertainty.
More detailed analysis has shown that the error in $\Omega_0$ prevails in the $A$ error.
Besides, published $\Omega_0$ results are not statistically consistent, unlike $R_0$ measurements.
So, more attention is needed to compute the best estimate of $\Omega_0$.

\clearpage

\begin{table}[ht!]
\centering
\def\arraystretch{0.75}
\begin{tabular}{lll}
\hline
\multicolumn{1}{c}{$R_0$} & \multicolumn{1}{c}{$\sigma$} & \multicolumn{1}{c}{Reference} \\
\hline
8.3    & 0.3  & Gerasimenko, 2004 \\
7.7    & 0.15 & Babusiaux \& Gilmore, 2005 \\
8.01   & 0.44 & Avedisova, 2005 \\
8.7    & 0.6  & Groenewegen \& Blommaert, 2005 \\
7.2    & 0.3  & Bica, et al., 2006 \\
7.52   & 0.36 & Nishiyama, et al. 2006 \\
8.1    & 0.7  & Shen \& Zhu, 2007 \\
7.4    & 0.3  & Bobylev, et al., 2007 \\
7.94   & 0.45 & Groenewegen, et al., 2008 \\
8.16 * & 0.5  & Ghez, et al., 2008 \\
8.07 * & 0.35 & Trippe, et al. 2008 \\
8.33 * & 0.35 & Gillessen, et al., 2009 \\
8.7    & 0.5  & Vanhollebeke, et al., 2009 \\
7.58   & 0.40 & Dambis, 2009 \\
8.4  * & 0.6  & Reid, et al., 2009 \\
7.75   & 0.5  & Majaess, et al., 2009 \\
8.24   & 0.43 & Matsunaga, et al., 2009 \\
7.9  * & 0.75 & Reid, et al., 2009 \\
7.7    & 0.4  & Dambis, 2010 \\
8.1    & 0.6  & Majaess, 2010 \\
8.3  * & 1.1  & Sato, et al., 2010 \\
7.80 * & 0.26 & Ando, et al., 2011 \\
8.29   & 0.16 & McMillan, 2011 \\
7.9    & 0.36 & Matsunaga, et al., 2011 \\
8.03   & 0.70 & Liu \& Zhu, 2011 \\
8.54   & 0.42 & Pietrukowicz, et al., 2012 \\
7.7  * & 0.4  & Morris, et al., 2012 \\
8.27   & 0.29 & Schoenrich, 2012 \\
8.05 * & 0.45 & Honma, et al., 2012 \\
7.51   & 0.23 & Bobylev, 2013 \\
8.24   & 0.43 & Matsunaga, et al., 2013 \\
8.38 * & 0.18 & Reid, 2013 \\
8.08   & 0.44 & Zhu \& Shen, 2013 \\
8.2    & 0.35 & Nataf, et al., 2013 \\
7.4    & 0.21 & Francis \& Anderson, 2013 \\
\hline
\end{tabular}
\hspace{1ex}
\begin{tabular}{lll}
\hline
\multicolumn{1}{c}{$\Omega_0$} & \multicolumn{1}{c}{$\sigma$} & \multicolumn{1}{c}{Reference} \\
\hline
27.6  & 1.7  & Bedin, et al., 2003 \\
32.8  & 1.2  & Olling \& Denhen, 2003 \\
25.3  & 2.6  & Kalirai, et al., 2004 \\
28.0  & 0.6  & Bobylev, 2004 \\
29.45 & 0.15 & Reid \& Brunthaler, 2004 \\
29.96 & 1.29 & Zhu, 2006 \\
26.0  & 0.3  & Bobylev et al., 2007 \\
30.7  & 1.0  & Lepine, et al., 2008 \\
27.67 & 0.61 & Bobylev, et al., 2008 \\
28.06 & 1.04 & Ghez, et al., 2008 \\
30.2  & 1.0  & Dambis, 2009 \\
30.3  & 0.9  & Reid, et al., 2009a \\
29.8  & 1.0  & Bovy, et al., 2009 \\
31    & 1    & Melnik \& Dambis, 2009 \\
27.27 & 1.04 & Dambis, 2010 \\
30.65 & 0.85 & Macmillan \& Binney, 2010 \\
31.0  & 1.2  & Bobylev \& Bajkova, 2010 \\
27.3  & 0.8  & Ando, et al., 2011 \\
28.7  & 1.3  & Nagayama, et al., 2011 \\
30.4  & 1.5  & Stepanishchev \& Bobylev, 2011 \\
31.5  & 0.9  & Bobylev \& Bajkova, 2011 \\
29.27 & 1.04 & Liu \& Zhu, 2011 \\
28.8  & 0.8  & Bajkova \& Bobylev, 2012 \\
27.5  & 0.5  & Bobylev \& Bajkova, 2012 \\
28.78 & 1.04 & Schoenrich, 2012 \\
31.09 & 0.78 & Honma, et al., 2012 \\
31.63 & 3.31 & Bobylev, 2013 \\
28    & 2    & Nagayama, et al., 2013 \\
29.0  & 1.0  & Reid, 2013  \\
32.38 & 1.04 & Bobylev \& Baikova, 2013 \\
\hline
&&\\
&&\\
&&\\
&&\\
&&\\
&&\\
\end{tabular}
\caption{$R_0$ [kpc] and $\Omega_0$ [km s$^{-1}$ kpc$^{-1}$] estimates. Direct $R_0$ measurements are marked with asterisk.}
\label{tab:data}
\end{table}

\begin{table}[ht!]
\centering
\begin{tabular}{ccccc}
\hline
Interval & $R_0$ data & $R_0$ & $\Omega_0$ & $A$ \\
\hline
2008--2013 & all    & $8.06 \pm 0.12$ & $29.59 \pm 0.75$ & $4.96 \pm 0.26$ \\
2008--2013 & direct & $8.14 \pm 0.15$ & $29.59 \pm 0.75$ & $5.01 \pm 0.27$ \\
2003--2013 & all    & $8.00 \pm 0.14$ & $29.28 \pm 0.66$ & $4.83 \pm 0.24$ \\
\hline
\end{tabular}
\caption{Results of computation of $R_0$ [kpc], $\Omega_0$ [km s$^{-1}$ kpc$^{-1}$], and $A$ [$\mu$as/yr.].}
\label{tab:result}
\end{table}

\vspace*{0.3cm}

\noindent {\large 3. CONCLUSION}

\smallskip

We derived the current best estimate of the GA constant using all available measurements of the Galactic rotation parameters made during last 5 years,
which yields the result $A = 4.96 \pm 0.26$~$\mu$as/yr.
For practical applications we suggest to use the value $A$ = 5~$\mu$as/yr.
Using this value of the GA constant allows one to eliminate about 90\% of the GA effect.
Remaining uncertainty in proper motion of about 0.5~$\mu$as/yr is negligible nowadays.
Thus the proposed value of the GA constant can be safely used in practice during coming years, presumably for at least the nearest decade,
until new VLBI and space observations provide substantially better result.

\bigskip
\noindent {\it Acknowledgements.} The author is grateful to the organizers of the conference for the travel support.

\vspace*{0.7cm}

\noindent {\large 4. REFERENCES}

{

\leftskip=5mm
\parindent=-5mm

\smallskip




Malkin, Z.M., 2011, Astron. Rep. 55, pp. 810--815.

Malkin, Z., 2012, arXiv:1202.6128.

Malkin, Z.M., 2013, Astron. Rep. 57, 882--887.

}

\end{document}